\documentclass[prd,twocolumn,a4paper,superscriptaddress,floatfix]{revtex4}
\usepackage{graphicx}
\begin{document}
\newcommand{\be}{\begin{equation}}
\newcommand{\ee}{\end{equation}}
\newcommand{\bq}{\begin{eqnarray}}
\newcommand{\eq}{\end{eqnarray}}
\newcommand{\bsq}{\begin{subequations}}
\newcommand{\esq}{\end{subequations}}
\newcommand{\bc}{\begin{center}}
\newcommand{\ec}{\end{center}}
\newcommand\lapp{\mathrel{\rlap{\lower4pt\hbox{\hskip1pt$\sim$}} \raise1pt\hbox{$<$}}}
\newcommand\gapp{\mathrel{\rlap{\lower4pt\hbox{\hskip1pt$\sim$}} \raise1pt\hbox{$>$}}}
\def\E{{\cal E}}

\title{Evolution of Hybrid Defect Networks}
\author{C.J.A.P. Martins}
\email[Electronic address: ]{Carlos.Martins@astro.up.pt}
\affiliation{Centro de Astrof\'{\i}sica, Universidade do Porto, Rua das Estrelas, 4150-762 Porto, Portugal}
\affiliation{DAMTP, University of Cambridge, Wilberforce Road, Cambridge CB3 0WA, United Kingdom}
\date{14 September 2009}
\begin{abstract}
We apply a recently developed analytic model for the evolution of monopole networks to the case of monopoles attached to one string, usually known as hybrid networks. We discuss scaling solutions for both local and global hybrid networks, and also find an interesting application for the case of vortons. Our quantitative results agree with previous estimates in indicating that the hybrid networks will usually annihilate soon after the string-forming phase transition. However, we also show that in some specific circumstances these networks can survive considerably more than a Hubble time.
\end{abstract}
\maketitle

\section{\label{intro}Introduction}

The pioneering work of Kibble \cite{KIBBLE} has shown that topological defects necessarily form at cosmological phase transitions. What type of defect network is formed and its basic properties (such as whether or not it is long-lived) will depend on the characteristics of each phase transition, and most notably on the specific symmetry being broken. Understanding these processes, as well as the subsequent evolution of these networks, is a key aspect of particle cosmology. A thorough overview of the subject can be found in the book by Vilenkin and Shellard \cite{VSH}.

Most of the work on defects in the past three decades has focused on cosmic strings. This is justified on the grounds that they are usually cosmologically benign, and are a generic prediction of inflationary models based on Grand Unified Theories \cite{Rachel1,Rachel2} or branes \cite{Branes1,Branes2}. By contrast domain walls and monopoles are cosmologically dangerous (and any models in which they arise are strongly constrained). However, a solid understanding of the latter is still important, as it is the only way one can trust these constraints. Moreover, it is becoming increasingly clear, particularly in the context of models with extra dimensions such as brane inflation, that hybrid defect networks will often be produced. Two examples that have attracted considerable interest are semilocal strings and cosmic necklaces \cite{Dasgupta1,Chen,Dasgupta2}. 

This is the second report on an ongoing project which is addressing some of these issues. In a previous work \cite{MONOPOLES} we have developed an analytic model for the evolution of networks of local and global monopoles \cite{DIRAC,THOOFT,POLYAK}. The model is analogous to the velocity-dependent one-scale model for cosmic strings \cite{MS0,MS2}, which has been extensively tested against field theory \cite{ABELIAN,MS3} and Goto-Nambu simulations \cite{MS3,MS4}. In \cite{MONOPOLES} we have also discussed the solutions of this analytic model, and compared them with existing (relatively low-resolution) simulations of global monopoles. Here, after a brief overview of hybrid networks, we extend our analysis to the case of monopoles attached to one string (the so-called hybrid networks), and we also briefly discuss how to apply it to vortons.

\section{\label{review}Overview of Hybrid Networks}

We start with a brief overview of previous results on the evolution of hybrid networks, emphasizing the dynamical aspects we seek to model. A more detailed can be found in the Vilenkin and Shellard textbook \cite{VSH} and in other references that will be pointed out where appropriate. We will first discuss the case of local symmetries, and then highlight relevant differences in the global case.

\subsection{Local case}

Hybrid networks of monopoles connected by strings will be produced, for example, in the following symmetry breaking scheme \cite{VILENKIN}
\be
G\to K\times U(1)\to K\,.
\ee
The first phase transition will lead to the formation of monopoles, and the second produces strings connecting monopole-antimonopole pairs; the corresponding defect masses will be
\be
m\sim \frac{4\pi}{e}\eta_m\label{mmass}
\ee
and
\be
\mu\sim 2\pi\eta^2_s\,.\label{smass}
\ee
The characteristic monopole radius and string thickness are
\be
\delta_m\sim (e\eta_m)^{-1}
\ee
\be
\delta_s\sim (e\eta_s)^{-1}\,,
\ee
In the simplest models all the (Abelian) magnetic flux of the monopoles is confined into the strings. In this case the monopoles are often dubbed 'beads'. However, in general (and in most realistic) models stable monopoles have unconfined non-Abelian magnetic charges. As in the case of isolated monopoles, the key difference between the two cases is that unconfined magnetic fluxes lead to Coulomb-type magnetic forces between the monopoles.

Up to the second transition the formalism that we have developed for plain monopoles \cite{MONOPOLES} obviously applies, but the hybrid network requires a separate treatment. From the point of view of analytic model-building, the hybrid case presents one crucial difference. In the case of isolated monopoles we saw that their evolution can be divided into a pre-capture and a post-capture period: captured monopoles effectively decouple from the rest of the network, and most of the radiative losses occur in the captured phase, where monopole-antimonopole pairs are bound and doomed to annihilation. This meant that the network's energy losses could be described as losses to bound pairs and we did not need to model radiative losses explicitly. Now compare this to the present scenario: the monopoles are now captured \textit{ab initio} (as soon as the strings form) and therefore we will need to rethink the loss terms---as well as to account for the force the strings exert on the monopoles.

The lifetime of a monopole-antimonopole pair is to a large extent determined by the time it takes to dissipate the energy stored in the string, $\epsilon_s\sim\mu L$ for a string of length $L$. This is because the energy in the string is typically {\bf larger} than (or at most comparable to) the energy of the monopole. Monopoles are pulled by the strings with a force $F_s\sim\mu\sim\eta^2_s$, while the frictional force acting on them is $F_{fri}\sim\theta  T^2 v$ (where $\theta$ is a parameter counting the number of degrees of freedom interacting with the monopoles). This friction term is of course already included in the evolution equations of the analytic model \cite{MONOPOLES,MS0,MS2}: the corresponding friction lengthscale is $\ell_f\equiv M/(\theta T^2)\sim\eta_m/(\theta T^2)$.

In a friction-dominated epoch the string tension should be compensated by the friction force, so the velocity of the monopoles can be estimated from
\be
m{\dot v}=\mu-\theta T^2v\sim0
\ee
which gives
\be
v\sim\frac{\mu}{\theta T^2}\,.
\ee
Note that in addition to the damping due to Hubble expansion and frictional scattering, we now have a third dynamical mechanism, which we may call string forcing. It's important to notice that in the radiation epoch the frictional scattering term has the same time dependence as the Hubble damping term, while in the matter era it is subdominant.

At string formation $F_s\sim F_{fri}$, so friction domination will end very quickly. This differs from the case of plain monopoles, whose evolution is always friction-dominated in the radiation era \cite{MONOPOLES}. At low temperatures the string tension is much greater than the friction force, and a monopole attached to a string moves with a proper acceleration $a=\mu/m\sim\eta^2_s/\eta_m$.

If there are non-confined fluxes, accelerating monopoles can also lose energy by radiating gauge quanta. The rate of energy loss is expected to be given by the classical electromagnetism radiation formula 
\be
{\dot \epsilon_{gauge}}\sim -\frac{(ga)^2}{6\pi}\sim- \left(\frac{g\mu}{m}\right)^2\sim- \left(\frac{\mu}{\eta_m}\right)^2
\ee
where $g$ is the magnetic charge. This should be compared with the ratio of gravitational radiation losses
\be
{\dot \epsilon_{grav}}\sim - G\mu^2\,;
\ee
the ratio of the two is therefore
\be
\frac{\dot \epsilon_{grav}}{\dot \epsilon_{gauge}}\sim \left(\frac{\eta_m}{m_{Pl}}\right)^2\,,
\ee
so if there are unconfined fluxes the gauge radiation will be dominant (except in the extreme case where the monopoles form at the Planck scale itself). The characteristic timescales for this process to act on a monopole-antimonopole pair attached to a string of length $L$ can therefore be written
\be
\tau_{rad}\sim \frac{L}{Q}
\ee
where $Q_{gauge}=(\eta_s/\eta_m)^2$ for gauge radiation and $Q_{grav}=(\eta_s/m_{Pl})^2$ for gravitational radiation.

These radiation losses should be included in the model's evolution equation for the characteristic lengthscale $L$ of the monopoles. It's easy to see that the corresponding term has the form
\be
3\frac{dL}{dt}=-L\frac{\dot\rho}{\rho}\sim-L\frac{\dot\epsilon}{\epsilon}\sim\frac{L}{\tau_{rad}}=Q\,.\label{losses}
\ee
Note that in principle these terms should be velocity-dependent (this is discussed in \cite{MS2}). However, we will soon see that in this case the monopoles will become ultra-relativistic ($v\sim1$) shortly after the strings form and therefore the velocity-dependence is not crucial for the analysis.

Another possible dynamical mechanism is that of string intercommuting and consequently the possibility of energy losses from the production of string loops. Its importance relative to other energy loss mechanisms is much harder to estimate than for ordinary strings, since were we're looking for an indirect effect on the monopole evolution, and probably one can only quantify it by using numerical simulations. What one can say is that, as in the standard case, it should lead to a term in the $L$ equation of the form \cite{MS2}
\be
\frac{dL}{dt}=c_{hyb}v
\ee
where the dimensionless coefficient $c_{hyb}$ need not have the same value (or indeed the same order of magnitude) as the standard one. We can now observe that since one expects to be dealing with ultra-relativistic string ($v\sim1$) the velocity dependence will again not be crucial, so we can group this term with that coming from radiation losses (Eq. \ref{losses}) and replace the coefficient $Q$ with an effective $Q_\star$ which will include the effects of gauge radiation (if it exists), gravitational radiation and loop production.

It is expected that monopole-antimonopole pairs will annihilate very quickly once the friction force becomes unimportant. The characteristic monopole velocity can be estimated to be approximately
\be
v\sim\left(\frac{\mu L}{m}\right)^{1/2},
\ee
or $v\sim1$ if the above is larger than unity. In the wake of the above discussion the corresponding rate of energy loss is naively estimated to be ${\dot\epsilon}\sim-T^2v^2$. The approximate lifetime of a pair should then be
\be
\tau\sim\frac{\mu L}{T^2v^2}\,,
\ee
which in the non-relativistic case can be simply written
\be
\tau_{nr}\sim\frac{m}{T^2}\,,
\ee
while for the ultra-relativistic case
\be
\tau_{rel}\sim\frac{\eta_s}{T^2}\,.
\ee
In either case monopoles annihilate in a timescale shorter than a Hubble time, which in these units is
\be
t\sim\frac{m_{Pl}}{T^2}\,.
\ee

\subsection{Global case}

One can also have hybrid networks of global monopoles connected by global strings \cite{VILENKIN}. Just as in the case of plain monopoles \cite{MONOPOLES}, the scale-dependent monopole mass and the different behavior of the forces between monopoles in this case (long-range rather than Coulomb-type) changes the detailed properties of these defects and warrants a separate treatment. 

The tension of a global string is
\be
F_s\sim 2\pi\eta^2_s\ln{\frac{L}{\delta_s}}\,,
\ee
so now there's an additional logarithmic correction, while the long-range force between the monopoles is
\be
F_m\sim 4\pi \eta^2_m\,.
\ee
If $\eta_m >> \eta_s$ the monopoles initially evolve as if they were free, with $L\sim t$. Note that this scaling law implies that there are some monopole annihilations. The strings become dynamically important when $F_s\sim F_m$, that is
\be
\ln{\left(t \eta_s\right)} \sim \frac{\eta^2_m}{\eta^2_s}\,,\label{globstrd}
\ee
at which point they pull the monopole-antimonopole pairs together, and the network is expected to disappear within a Hubble time.

The scale-dependent monopole mass is important for the friction lengthscale, which is now $\ell_f\equiv M/(\theta T^2)\sim\eta_m^2 L/(\theta T^2)$. Another difference is that instead of gauge radiation we now have Goldstone boson radiation, whose rate of energy loss is
\be
{\dot \epsilon_{gold}}\sim - \eta_m^2\,;
\ee
notice that this is much stronger than the energy loss rates due to gauge radiation (except if the monopoles and strings form at the same energy scale, in which case they will be comparable) and due to gravitational radiation (except if both form that the Planck scale, in which case they will all be comparable).

This energy loss term should be similarly included in the evolution equation for the characteristic lengthscale $L$. Note that there's a crucial difference between this and the gauge case: here the string energy is typically {\bf smaller} than the monopole energy, $E_m\sim\eta^2_mL$. The string is therefore (to some extent) irrelevant, and in this case the $Q$ coefficient is $Q_{gold}\sim cv$ as in the case of isolated global monopoles \cite{MONOPOLES}. Notice the presence of the velocity-dependence, although one expects that the velocities will always be relativistic. The effect of string loop production can in principle be accounted for by a redefined (effective) coefficient, $c_\star$.

\section{\label{models}Modeling Hybrids}

The above overview leads to the expectation that the network will annihilate shortly after the strings form. In fact we will see below that there are circumstances where the network can survive considerably more than a Hubble time, although it has to be said that we do expect this to be the exception rather than the rule.

We will therefore proceed fairly quickly, discussing only the late-time regime where the force due to the strings is dominating and the hybrid networks are therefore about to annihilate. As discussed in Sect. \ref{review}, in the local case this will happen very shortly after the string-forming phase transition. In the global case, however, there will be an intermediate epoch where the monopoles evolve as if free (because the strings are comparatively very light), and the strings only become dynamically important at an epoch given by Eq. (\ref{globstrd}).

\subsection{Local case}

In this case our evolution equations for the characteristic lengthscale $L$ and RMS velocity $v$ of the monopoles take the form
\be
3\frac{dL}{dt}=3HL+v^2L\left(H+\frac{\theta T^2}{\eta_m}\right)+Q_\star
\ee
where $Q_\star$ includes the energy loss terms from gauge radiation (if it exists), gravitational radiation and loop production discussed earlier, possibly with some coefficient of order unity, and
\be
\frac{dv}{dt}=(1-v^2)\left[\frac{k_m}{\eta_mL^2}\frac{L}{d_H}+k_s\frac{\eta_s^2}{\eta_m}-v\left(H+\frac{\theta T^2}{\eta_m}\right)\right]\,.
\ee
The velocity equation now has two accelerating terms, due to the strings and the inter-monopole Coulomb forces, which we parametrize with coefficients $k_s$ and $k_m$ which are expected to be of order unity. An exception is the Abelian case, where there are no Coulomb forces, so $k_m=0$ in this case.

It's important to realize that friction will play a crucial role in the radiation era, where it is more important than the Hubble damping term (the opposite is true for the matter era). Indeed in the radiation era we can write
\be
\frac{1}{\ell_d}=H+\theta\frac{T^2}{\eta_m}=\frac{1}{t}\left(\frac{1}{2}+\theta\frac{m_{Pl}}{\eta_m}\right)\equiv\frac{\lambda_\star}{t}
\ee
where in the last step we have defined an effective coefficient that is usually much larger than unity (except if there were no particles interacting with the string, $\theta=0$). On the other hand, in other cosmological epochs (in particular the matter-dominated era) the friction term is negligible, and we have $\lambda_\star=\lambda$ (where we are defining $a\propto t^\lambda$) as usual.

It is illuminating to start by comparing the various terms in the velocity equation. The ratio of the monopole and string accelerating terms is
\be
\frac{F_m}{F_s}=\frac{k_m}{k_s}\frac{1}{Ld_H\eta^2_s}\sim\left(\frac{\delta_s}{L}\right)\left(\frac{\delta_m}{d_H}\right)
\ee
which is always much less than unity (except if $k_s$ happened to be extremely small). The Coulomb forces are always negligible relative to the string forces. This was expected, since as we pointed out earlier the energy in the strings is typically larger than that in the monopoles. A useful consequence is that we do not need to treat the Abelian and non-Abelian cases separately.

Now let us compare the damping and string acceleration terms
\be
\frac{F_d}{F_s}=\frac{\lambda_\star}{k_s}\left(\frac{\eta_m}{m_{Pl}}\right)\left(\frac{T}{\eta_s}\right)^2v\sim\frac{\theta}{k_s}\left(\frac{T}{\eta_s}\right)^2v
\ee
where in the last step we assumed that we are in the radiation era. Given that the evolution of the monopoles before string formation leads to a scaling law $v\propto a^{-1}$ for monopole velocities \cite{MONOPOLES}, which are therefore extremely small when the strings form, the above ratio is always less than unity.

This analysis therefore quantitatively confirms the naive expectation that as soon as the strings are formed the string acceleration term will dominate the dynamics and drive the monopole velocity to unity. Recalling that the initial monopole velocity can effectively be taken to be zero, we can write the following approximate solution of the monopole velocity equation
\be
\ln{\frac{1+v}{1-v}}=2f_s(t-t_s)
\ee
and we can compare the epoch at which the monopoles become relativistic ($t_c$) with that of the string-forming phase-transition ($t_s$), finding
\be
\frac{t_c}{t_s}\sim1+\frac{\eta_m}{m_{Pl}}\,, \label{tdecayv}
\ee
so they become relativistic less than a Hubble time after the epoch of string formation; notice that this ratio depends only on the energy scale of the monopoles, and not on that of strings.

We can now proceed to look for solutions for the characteristic monopole lengthscale $L$, assuming for simplicity that $v=1$ throughout. We can look for solutions of the form $L\propto t^\alpha$ for generic expansion rates ($a\propto t^{\lambda}$), and it is straightforward to see that there are two possibilities. There is a standard linear scaling solution
\be
L=\frac{Q_\star}{3(1-\lambda)-\lambda_\star}t
\ee
in which the energy loss terms effectively dominate the dynamics. In this case the monopole density decays slowly relative to the background density,
\be
\frac{\rho_m}{\rho_b}\propto t^{-1}\,.
\ee
This scaling law can in principle exist for any cosmological epoch provided $\lambda<3/4$ (for example, in the matter era we have $L=3Q_\star t$). However, in the radiation epoch we would need the unrealistic $\theta m_{Pl}/\eta_m<2$, which effectively would mean that friction is absent ($\theta=0$).

The alternative scaling solution, for epochs when friction dominates over Hubble damping (such as the radiation era) and also for any epoch with $\lambda\ge3/4$ (even without friction) has $L$ growing with a power $\alpha>1$ given by
\be
\alpha=\lambda+\frac{1}{3}\lambda_\star=\frac{1}{3}\left(4\lambda+\theta\frac{m_{Pl}}{\eta_m}\right)\,;
\ee
if there is no friction the scaling law can simply be written
\be
L\propto a^{4/3}\,,
\ee
from which one sees that although the Hubble term is dominant, the scaling is faster than conformal stretching ($L\propto a$) because the velocities are ultra-relativistic. The important point here is that in this regime $L$ grows faster than $L\propto t$, and the number of monopoles per Hubble volume correspondingly decreases.

At the phenomenological level of our one-scale model, this corresponds to the annihilation and disappearance of the monopole network. We can easily estimate the timescale for this annihilation---it will occur when the number of monopoles per Hubble volume drops below unity (or equivalently $L>d_H$). Assuming a lengthscale $L_s=st_s$ at the epoch of string formation (note that the evolution of the monopole network before the stings form is such that $s$ can be much smaller than unity), we easily find
\be
\frac{t_a}{t_s}\sim1+\frac{3}{\theta}\left(\frac{2}{s}-1\right)\frac{\eta_m}{m_{Pl}}\,, \label{tdecayr}
\ee
which is comparable to the estimate we obtained using the velocity equation, Eq. (\ref{tdecayv}). Notice that this is a much faster timescale than the one associated with radiative losses, which can therefore be consistently neglected in this case. There's nothing unphysical about this 'superluminal' behaviour, as explained in \cite{EVERETT}. We do have a physical constraint that the timescale for the monopole disappearance should not be smaller than the (initial) length of the string segments, $t_a\ge L_s$, but this should always be the case for the above solutions.

This analysis shows that the monopoles must annihilate during the radiation epoch, if one wants to solve the monopole problem by invoking nothing but a subsequent string-forming phase transition. Any monopoles that survived into the matter era would probably be around today. It also shows that monopoles can also be diluted by a sufficiently fast expansion period. Inflation is of course a trivial example of this, but even a slower expansion rate $3/4<\lambda<1$ would be sufficient, provided it is long enough to prevent the monopoles from coming back inside the horizon by the present time.

\subsection{An aside: string evolution}

Notice that in the above we haven't yet said anything about the evolution of the strings. One could try to define a string characteristic lengthscale in the usual way, $\rho_s=\mu_s/L^2_s$. However, one should be careful about doing this, since in this case there is no a priori expectation that the distribution of string segments connecting the pairs of monopoles and antimonopoles will form an effectively Brownian network. (In fact it is quite likely that it doesn't, although this is something that warrants numerical investigation.)

A safer and simpler way of describing them is to look at the evolution of an individual string segment, and then make use of the fact that we already know how the monopole number and characteristic lengthscale evolve. The evolution of a given segment of a local string of length $\ell$ in the context of the VOS model has been studied in \cite{MS1}. In the present notation the evolution equation has the form
\be
\frac{d\ell}{dt}=(1-2v^2)H\ell-\frac{\ell}{\ell_f}v^2-Q_\star\,;
\ee
note that as expected energy loss mechanisms reduce the length of the string segment. There's an analogous equation for the string segment's velocity, but we already know we can safely assume $v\sim1$. Using this and the specific form of the friction term, the above equation can in fact be written
\be
\frac{d\ell}{dt}=-\frac{\lambda_\star}{t}\ell-Q_\star\, \label{dynforsegments}
\ee
which can easily be integrated to yield (for $\ell=\ell_s=st_s$ at the string formation epoch $t=t_s$)
\be
\ell(t)=\ell_s\left(\frac{t_s}{t}\right)^{\lambda_\star}+\frac{Q_\star}{1+\lambda_\star}\left[t_s\left(\frac{t_s}{t}\right)^{\lambda_\star}-t\right]\,.\label{timedecay0}
\ee
We emphasize that for simplicity we are assuming that all string segments are formed with the same length: at a detailed level this is unrealistic, but it is nevertheless sufficient to provide reliable qualitative estimates.

We can now estimate the monopole annihilation epoch by simply looking for $\ell\sim0$. As before, the answer will depend on the relative importance of the the radiative loss and friction terms. If friction is negligible and the radiative losses dominate, then we find the radiative decay timescale 
\be
\frac{t_r}{t_s}\sim\left(1+s\frac{1+\lambda_\star}{Q_\star}\right)^{1/1+\lambda_\star}\,,\label{timedecay1}
\ee
which is much larger than unity. On the other hand, if friction is important and the $Q$ term is subdominant the annihilation timescale is the much faster
\be
\frac{t_a}{t_s}\sim1+\frac{1}{\theta}\frac{\eta_m}{m_{Pl}}\,,\label{timedecay2}
\ee
which is again comparable to our previous estimates, given by Eqs. (\ref{tdecayv}) and (\ref{tdecayr}).

Since each piece of string is connecting two monopoles, a very simple estimate of the total length in string in a Hubble volume (and hence of the string density) is obtained multiplying half the number of monopoles in that volume by the typical length of each segment. This leads to
\be
\frac{\rho_s}{\rho_m}\propto \frac{\eta^2_s}{\eta_m}\ell(t)
\ee
and as expected the string density decays relative to that of the monopoles.

\subsection{Global case}

In the global case the evolution equations will be \cite{MONOPOLES}
\be
3\frac{dL}{dt}=3HL+v^2\frac{L}{\ell_d}+c_\star v
\ee
and
\be
\frac{dv}{dt}=(1-v^2)\left[\frac{k_m}{L}\left(\frac{L}{d_H}\right)^{3/2}+\frac{k_s}{L}\frac{\eta^2_s}{\eta^2_m}\ln{\frac{L}{\delta_s}}-\frac{v}{\ell_d}\right]\,,
\ee
and the scale dependence of the monopole mass and string tension imply that this case differs in two ways from the local case.

Firstly, the damping term at the string formation epoch (assumed to be in the radiation epoch) can be written
\be
\frac{1}{\ell_d}=H+\theta\frac{T^2}{\eta_m^2L}\sim\frac{1}{t}\left[\frac{1}{2}+\theta\left(\frac{\eta_s}{\eta_m}\right)^2\right]\,.
\ee
The friction term is now subdominant relative to Hubble damping, and of course it decreases faster than it. Friction can therefore be ignored in the analysis.

Secondly, the monopole acceleration due to the (global) strings now has an extra logarithmic correction, but more importantly the logarithmically divergent monopole mass again implies that this force is inversely proportional to $L$ (as opposed to being constant in the local case). The large lengthscale inside the logarithm should be the string lengthscale, but we have substituted it for the monopole one, which we'll simple denote $L$.

Starting again with the velocity equation, the ratio of the string and monopole acceleration terms is now
\be
\frac{F_s}{F_m}\sim\frac{k_s}{k_m}\frac{\eta^2_s}{\eta^2_m}\frac{d_H^{3/2}}{L^{3/2}}\ln{(L\eta_s)}\,,
\ee
and in particular at the epoch of string formation we have
\be
\left(\frac{F_s}{F_m}\right)_{t_s}\sim\frac{k_s}{k_m}\frac{\eta^2_s}{\eta^2_m}\ln{\frac{m_{Pl}}{\eta_s}}\,,
\ee
so now $F_s$ is initially sub-dominant, except if we happen to have $\eta_s\sim\eta_m$. Moreover, given that $L\propto t$ with a proportionality factor not much smaller than unity, while monopoles evolve freely (before the effect of the strings is important), the ratio will only grow logarithmically, and so the effect of the strings might not be felt for a very long time. Specifically this should happen at an epoch
\be
\frac{t_r}{t_s}\sim\frac{\eta_s}{m_{Pl}}exp\left(\frac{\eta^2_m}{\eta^2_s}\right)\,;
\ee
naturally this is exactly the same as Eqn. (\ref{globstrd}).

While strings are dynamically unimportant we have $v=const.$ as for free global monopoles, and even when they become important the velocity will still grow very slowly (logarithmically) towards unity. Hence, although the ultimate asymptotic result is the same in both cases ($v\to1$), the timescale involved should be much larger in the global case.

Moreover, recall from \cite{MONOPOLES} that in the gauge case the monopole velocities before string formation were necessarily non-relativistic and indeed extremely small, and it is the strings that make them reach relativistic speeds. In the global case this is usually not so, as the monopoles will typically have significant velocities while they are free (although their magnitude depends on model parameters that need to be determined numerically). We can therefore say that the impact of the forces due to the strings is vastly smaller in the global case. All this is due to the fact that the force due to the strings is inversely proportional to $L$ instead of being constant.

As for the evolution of $L$, at early times (before strings become important) we have $L\propto t$ just like for free global monopoles. Eventually the strings push the monopole velocities close(r) to unity, and in this limit the $L$ evolution equation looks just like that for the global case with the particular choices $\lambda_\star=\lambda$ and $Q_\star=c$. However, we must be careful about timescales, since here the approach to $v=1$ is only asymptotic.

Bearing this in mind, for $\lambda<3/4$ we will still have linear scaling solution
\be
L=\frac{c}{3-4\lambda}t\,;
\ee
notice that this is exactly the ultra-relativistic ($v=1$) scaling solution we have described in \cite{MONOPOLES}---cf. Eqs. (70-71) therein. Therefore, if the network happened to be evolving in the other (subluminal) linear scaling solution, the only role of the strings would be to gradually switch the evolution to the ultra-relativistic branch. This scaling law can in principle occur both in the radiation and in the matter eras, and it follows that in this case the monopoles will not disappear at all, but will continue to scale indefinitely (with a constant number per Hubble volume).

In this case an analysis in terms of the length of each string segment would have to take into account an initial distribution of lengths. Moreover, since the initial velocities need not be ultrarelativistc, the larger segments (which will have smaller coherent velocities) should grow at early times, while the smaller ones will shrink. The decay time will obviously depend on the initial size. We note that such a behavior has been seen in numerical simulations of semilocal strings \cite{SEMILOCALSIM}.

The alternative scaling solution, for any epoch with $\lambda\ge3/4$ has $L$ growing with as
\be
L\propto a^{4/3}\,,
\ee
and again corresponds to the annihilation and disappearance of the monopole network, which would occur at
\be
\frac{t_a}{t_r}\sim\frac{1}{\left[(1-\lambda)r\right]^{3/(4\lambda-3)}}\,
\ee
for an initial lengthscale $L_s=rt_r$; given that we now expect $r$ to be not much smaller than unity, this is likely to be very soon after the onset of this scaling regime---that is, very soon after the velocities become ultra-relativistic. (Notice that in the local case the monopoles became ultra-relativistic very soon after the strings formed so $t_c\sim t_s$, but in the present case $t_c>>t_s$, except if the free monopoles were already evolving in the ultra-relativistic branch.)

\section{\label{topology}Topological Stability}

There is in principle a further energy loss mechanism that we have not discussed so far. These strings are not topologically stable: they can break, producing a monopole-antimonopole pair at the new ends. This process was first discussed in \cite{VILENKIN} and more recently in \cite{LEBLOND}. 

This is a tunneling process (usually called the Schwinger process), and its probability per unit string length and per unit time has been estimated to be \cite{VILENKIN,Monin1,Monin2}
\be
P\sim \frac{\mu}{2\pi}\exp{\left(-\pi\frac{m^2}{\mu}\right)}\,.
\ee
By substituting Eqs.(\ref{mmass}-\ref{smass}) one notices that in typical GUT models the exponent is very large, say $\sim10^3(\eta_m/\eta_s)^2$. This implies that this probability is negligibly small, even in the most favorable case where the strings and monopoles form at the same energy scale. Therefore, for all practical purposes such strings (whether they are local or global) can be considered stable.

Nevertheless, it is conceivable that in string-inspired models such as brane inflation there are models (or regions of parameter space in some models) where the exponent is not much larger than unity. Here we briefly discuss what happens in this case. A different scenario, where the bead- and string-forming phase transitions are separated by an inflationary epoch (and thus the beads are initially outside the horizon), has been recently discussed in \cite{LEBLOND}.

Again it is simplest to look at the effect of this additional energy loss term on an individual string segment. It's straightforward to show that Eq. (\ref{dynforsegments}) now becomes
\be
\frac{d\ell}{dt}=-\frac{\lambda_\star}{t}\ell-Q_\star-P\ell^2\,. \label{dynbreaking}
\ee
Assuming that the exponent in the Schwinger term is of order unity, it's easy to see that for large segments (that is those much larger than the string thickness $\delta_s$) the Schwinger term is the dominant one, and it quickly breaks up the segment into smaller ones. On average (and neglecting the standard damping and energy loss terms) the length of a given segment is expected to shrink as
\be
\ell(t)=\frac{\ell_s}{1+P\ell_s(t-t_s)}\,
\ee
and the epoch (relative to that of string formation) at which this process would, on average, have reduced the segment to a size $\delta_s$ is
\be
\frac{t_\delta}{t_s}=1+\frac{1}{e}\frac{\eta_s}{m_{Pl}}
\ee
which is typically less than a Hubble time. As the segments shrink the standard damping and especially the radiative energy loss terms will become more important, and we would eventually switch to a solution of the type discussed earlier, cf. Eq. (\ref{timedecay0}). The segment size when this happens will be very small, so $s<<1$ in Eq. (\ref{timedecay1}), and these segments should therefore decay in much less than a Hubble time.

\section{\label{aside}A digression: vortons}

Since vortons \cite{VORTONS} are to a large extent point-like objects decoupled from the string network that produced them, it would be interesting to use the monopole model to describe the evolution of their overall density.

The most naive approach would be simply to say that there are no inter-vorton or other forces (other than friction and Hubble damping) affecting their dynamics. This immediately implies that there will be no acceleration term in the velocity equation ($k=0$), and hence vortons will necessarily be non-relativistic.

At the same naive level there should be no energy losses to account for in the lengthscale evolution equation. Classically this is justifiable by saying that we can assume that all vortons form around the same epoch (related to the superconducting phase transition, which need not be the same as the string-forming phase transition) and that once formed a vorton stays there forever (or at least has a very long lifetime)---these issues are discussed in \cite{VSH}. Of course we're not addressing the issue of the vorton density at formation (which would set the initial conditions for the model). A different issue which we are also neglecting is that there could well be an energy loss term due to quantum effects.

With these caveats in mind, it follows from our previous work \cite{MONOPOLES} that the evolution equations will be
\be
3\frac{dL}{dt}=(3+v^2)HL+v^2\frac{L}{\ell_f}
\ee
and
\be
\frac{dv}{dt}=-(1-v^2)v\left(H+\frac{1}{\ell_f}\right)\,.
\ee
Note that here $L$ is a characteristic vorton lengthscale, which is related to the vorton density through
\be
\rho_v=\frac{M_v}{L^3}\sim\frac{\eta^2_s\ell_v}{L^3}
\ee
where $\eta_s$ is the string symmetry breaking scale ($\mu\sim\eta^2_s$) and $\ell_v$ is an effective vorton length (not necessarily its radius/size, since there will be a significant amount of energy associated with the charge/current). For simplicity (and consistently with the above approximations) we will further assume that this length is a constant.

The simplest possibility we can consider is the case where friction is negligible ($\ell_f\to\infty$). The analysis of this case is trivial, and it leads to
\be
L\propto a\,,\qquad v\propto\frac{1}{a}\propto\frac{1}{L}\,.
\ee
This of course leads to $\rho_v\propto a^{-3}$, which is in fact what has been found in previous analysis both for generic vortons \cite{BRANDENBERGER} and for the specific case of chiral vortons \cite{CARTERDAVIS}. In either case there are several possibilities for the initial density, depending on details of the underlying particle physics model. There is no published analysis of vorton scaling velocities, although the effect of superconducting currents on the loop velocities has been discussed in \cite{MSVORTONS}, and the above scaling law for velocities is well known for defects in condensed matter systems.

However it's also important to consider the role of friction, which in this context should not have the same form as the standard case. This is discussed in \cite{CHUDNOVSKY} and also in the Vilenkin and Shellard textbook \cite{VSH}. Because a superconducting string creates a magnetic field
\be
B_s(r)\sim2\frac{J}{r}
\ee
($J$ being the current) a string moving in a plasma is shielded by a magnetocylinder which contains the magnetic field which does not allow the plasma to penetrate. The magnetocylinder radius is
\be
r_s\sim\frac{J}{v\sqrt{\rho_p}}\,,
\ee
where $\rho_p$ is the plasma density, and this leads to a friction force per unit length
\be
F_p\sim Jv\sqrt{\rho_p}
\ee
which in our notation corresponds to a friction lengthscale
\be
\ell_f^{-1}\sim\frac{J\sqrt{\rho_p}}{\eta^2_s}\,.
\ee
Note that the current can have a maximum value
\be
J_{max}\sim e\sqrt{\mu}\sim \eta_s\,.
\ee

We can now study the effect of this friction term on the above evolution equations. It's easy to see that the scaling of the characteristic vorton scale is unchanged, in other words we still have $L\propto a$. However, there is an effect on the velocity equation, which now takes the form
\be
\frac{dv}{dt}=-(1-v^2)v\left(H+\frac{J\sqrt{\rho_p}}{\eta^2_s}\right)\,.
\ee
Now for a universe with a critical density
\be
\rho_p=\frac{3H^2}{8\pi G}
\ee
so in fact the two terms are proportional to the Hubble parameter, and which of the two dominates will depend on the strength of the current $J$. For a generic expansion law $a\propto t^\lambda$ we can write the scaling law for the vorton velocities as
\be
v\propto t^{-\beta}
\ee
where
\be
\beta=\lambda\left(1+\sqrt{\frac{3}{8\pi}}\frac{Jm_{Pl}}{\eta^2_s}\right)\,.
\ee
For small currents we have $v\propto t^{-\lambda}\propto a^{-1}$ as before, but the effect of the plasma is to reduce the velocities at a much faster rate. Note that for a maximal current the scaling exponent will be approximately $(m_{Pl}/\eta_s)$, which is much larger than unity.

It would be interesting to test this behavior numerically, as well as to understand the magnitude and form of energy loss terms due to quantum effects.

\section{\label{concl}Summary and outlook}

We have applied a recently developed analytic model for the evolution of monopole networks \cite{MONOPOLES} to the case hybrid networks of monopoles attached to a single string, and also to the case of vortons. We discussed scaling solutions for both local and global hybrid networks, generically confirming the expectation that the network will annihilate shortly after the strings form, but also highlighting the fact that there are circumstances where the network can be long-lived---although it has to be said that we do expect this to be the exception rather than the rule.

Studying the evolution of hybrid defects may seem relatively uninteresting because of their brief existence or even due to the topological instability discussed in Sect. \ref{topology}. However, this study has a second motivation. We have studied local strings attached to local monopoles and global strings attached to global monopoles. These are two simple test cases before tackling a third, more interesting case: local strings attached to global monopoles. This corresponds (modulo a few subtleties) to the case of semilocal strings \cite{SEMILOCAL1,SEMILOCAL2}, which we will address in a subsequent publication.

\begin{acknowledgments}
I am grateful to Ana Ach\'ucarro for valuable discussions and collaboration on related issues, and to Petja Salmi and Jon Urrestilla for many discussions in the early stages of this work. The work of C.M. is funded by a Ci\^encia2007 Research Contract, supported by FSE and POPH-QREN funds. I also thank the Galileo Galilei Institute for Theoretical Physics (where part of this work was done) for the hospitality and the INFN for partial support.
\end{acknowledgments}

\bibliography{hybrids}
\end{document}